\documentclass[aps,pra]{revtex4}
\usepackage{graphicx}

\begin{document}

\title{Deterministic secure direct communication by using swapping
quantum entanglement and local unitary operations \\
\thanks{*Email: zhangzj@wipm.ac.cn }}

\author{Z. J. Zhang and Z. X. Man  \\
{\normalsize Wuhan Institute of Physics and Mathematics, Chinese
Academy of Sciences, Wuhan 430071, China } \\
{\normalsize *Email: zhangzj@wipm.ac.cn }}

\date{\today}
\maketitle

\begin{minipage}{380pt}
A deterministic direct quantum communication protocol by using
swapping quantum entanglement and local unitary operations is
proposed in this paper. A set of ordered EPR pairs in one of the
four Bell states is used. For each pair, each of the two
legitimate users owns a photon of the entangled pair via quantum
channel. The pairs are divided into two types of group, i.e., the
checking groups and the encoding-decoding groups.  In the checking
groups, taking advantage of the swapping quantum entanglement and
Alice's (the message sender's) public announcement, the
eavesdropping can be detected provided that the number of the
checking groups is big enough. After insuring the security of the
quantum channel, Alice encodes her bits via the local unitary
operations on the encoding-decoding groups. Then she performs her
Bell measurements on her photons and publicly announces her
measurement results. After her announcement, the message receiver
Bob performs his Bell measurements on his photons and directly
extracts the encoding bits by using the property of the quantum
entanglement swapping. The security of the present scheme is also
discussed: under the attack scenarios to our best
knowledge, the scheme is secure. \\

PACS Number(s): 03.67.Hk, 03.65.Ud\\
\end{minipage}

Quantum key distribution (QKD) is an ingenious application of
quantum mechanics, in which two remote legitimate users (Alice and
Bob) establish a shared secret key through the transmission of
quantum signals. Much attention has been focused on QKD after the
pioneering work of Bennett and Brassard published in 1984 [1].
Till now there have been many theoretical QKDs[2-20]. They can be
classified into two types, the nondeterministic one [2-14] and the
deterministic one [15-20]. The nondeterministic QKD can be used to
establish a shared secret key between Alice and Bob, consisting of
a sequence of random bits. This secret key can be used to encrypt
a message which is sent through a classical channel. In contrast,
in the deterministic QKD, the legitimate users can get results
deterministically provided that the quantum channel is not
disturbed. It is more attractive to establish a deterministic
secure direct communication protocol by taking advantage of the
deterministic QKDs. However, different from the deterministic
QKDs, the deterministic secure direct communication protocol is
more demanding on the security. Hence, only recently a number of
deterministic secure direct protocols have been proposed
[15-16,19]. In these protocols, the quantum entanglement plays
very important roles. It is well known that quantum entanglement
swapping [21-24] can entangle two quantum system which do not
interact with each other and it has ever played very important
roles in some nondeterministic QKDs[7,15]. In this paper, we
present a deterministic secure direct communication protocol by
using swapping quantum entanglement and local unitary operations.
This communication protocol can be used to transmit securely
either a secret key or a plain text message.

Let us first describe the quantum entanglement swapping simply.
Let $|0\rangle$ and $|1\rangle$ be the horizontal and vertical
polarization states of a photon, respectively. Then the four Bell
states, $|\Psi ^{\pm }\rangle =(|01\rangle \pm |10\rangle
)/\sqrt{2}$ and $|\Phi ^{\pm }\rangle =(|00\rangle \pm |11\rangle
)/\sqrt{2}$, are maximally entangled states in the two-photon
Hilbert space.  Let the initial state of two photon pairs be the
product of any two of the four Bell states, such as $|\Psi
_{12}^{+}\rangle$ and $|\Psi _{34}^{+}\rangle$, then after the
Bell measurements on the photon 1 and 3 pair and the photon 2 and
4 pair, since the following equation holds,
\begin{eqnarray}
|\Psi _{12}^{+}\rangle\otimes |\Psi
_{34}^{+}\rangle=\frac{1}{2}(|\Psi _{13}^{+}\rangle |\Psi
_{24}^{+}\rangle-|\Psi _{13}^{-}\rangle|\Psi
_{24}^{-}\rangle+|\Phi _{13}^{+}\rangle|\Phi
_{24}^{+}\rangle-|\Phi _{13}^{-}\rangle|\Phi _{24}^{-}\rangle),
\end{eqnarray}
the total initial state (i.e., $|\Psi _{12}^{+}\rangle \otimes
|\Psi _{34}^{+}\rangle$) is projected onto $|\eta _{1}\rangle
=|\Phi _{13}^{+}\rangle \otimes |\Phi _{24}^{+}\rangle , |\eta
_{2}\rangle = |\Phi _{13}^{-}\rangle \otimes |\Phi
_{24}^{-}\rangle , |\eta _{3}\rangle =|\Psi _{13}^{+}\rangle
\otimes |\Psi _{24}^{+}\rangle$ and $|\eta _{4}\rangle =|\Psi
_{13}^-\rangle \otimes |\Psi _{24}^- \rangle$ with equal
probability of $\frac{1}{4}$ for each. It is seen that previous
entanglements between photons 1 and 2, and 3 and 4, are now
swapped into the entanglements between photons 1 and 3, and 2 and
4.  Therefore, if $|\Phi _{13}^{+}\rangle$ is obtained by the Bell
measurements, $|\Phi _{24}^{+}\rangle$ should be gained
affirmatively by the Bell measurements; if $|\Phi _{13}^-\rangle$
is obtained, then $|\Phi _{24}^-\rangle$ is arrived at; and so on.
This means that for a known initial state  the Bell measurement
results after the quantum entanglement swapping are correlated. In
the above example $|\Psi _{12}^{+}\rangle \otimes |\Psi
_{34}^{+}\rangle$ is chosen as the initial state. In fact, similar
results can also be arrived at provided that other choices of the
initial states are given. As can be seen as follows:
\begin{eqnarray}
\left\{ \begin{array}{c}|\Psi _{12}^{+}\rangle\otimes |\Psi
_{34}^{-}\rangle=\frac{1}{2}(|\Psi _{13}^{+}\rangle |\Psi
_{24}^{-}\rangle-|\Psi _{13}^{-}\rangle|\Psi
_{24}^{+}\rangle-|\Phi
_{13}^{+}\rangle|\Phi _{24}^{-}\rangle+|\Phi _{13}^{-}\rangle|\Phi _{24}^{+}\rangle), \\
 |\Psi _{12}^{+}\rangle\otimes |\Phi _{34}^{+}\rangle=\frac{1}{2}(|\Psi
_{13}^{+}\rangle|\Phi _{24}^{+}\rangle-|\Psi _{13}^{-}\rangle|\Phi
_{24}^{-}\rangle+|\Phi
_{13}^{+}\rangle|\Psi _{24}^{+}\rangle-|\Phi _{13}^{-}\rangle|\Psi _{24}^{-}\rangle), \\
 |\Psi _{12}^{+}\rangle\otimes |\Phi _{34}^{-}\rangle=\frac{1}{2}(|\Psi
_{13}^{+}\rangle|\Phi _{24}^{-}\rangle-|\Psi _{13}^{-}\rangle|\Phi
_{24}^{+}\rangle-|\Phi _{13}^{+}\rangle|\Psi
_{24}^{-}\rangle+|\Phi _{13}^{-}\rangle|\Psi _{24}^{+}\rangle).
\end{array}\right .
\end{eqnarray}
By the way, for the above four known initial states the
correlation of the Bell measurement results after the quantum
entanglement swapping is very useful in our protocol in detecting
the eavesdropping. On the other hand, it should also be noted that
different results by the Bell measurements correspond to different
initial states for the above four known initial states. For
examples, when $|\Psi _{13}^{+}\rangle$ and $|\Psi
_{24}^{-}\rangle$ are obtained by the Bell measurements, the
initial state should be $|\Psi _{12}^{+}\rangle\otimes |\Psi
_{34}^{-}\rangle$; when $|\Phi _{13}^{+}\rangle$ and $|\Psi
_{24}^{-}\rangle$ are obtained by the Bell measurements, the
initial state should be $|\Psi _{12}^{+}\rangle\otimes |\Phi
_{34}^{-}\rangle$; and so on. Incidentally, this property is also
used in our communication protocol. In addition, it is easily
verified that, the four Bell states can be transformed into each
other by some unitary operations, which can be performed locally
with nonlocal effects. For examples: Let $u_0$,$u_1$,$u_2$,$u_3$
be in turn the unitary
operations $\left( \begin{array}{cc} 1 & 0 \\ 0 & 1%
\end{array}%
\right) ,\left(
\begin{array}{cc}
1 & 0 \\
0 & -1%
\end{array}%
\right), $ $\left(
\begin{array}{cc}
0 & 1 \\
1 & 0%
\end{array}%
\right) ,\left(
\begin{array}{cc}
0 & -1 \\
1 & 0%
\end{array}%
\right) $ respectively, then $|\Psi _{34}^{+}\rangle$ will be in
turn transformed into $|\Psi _{34}^{+}\rangle$, $|\Psi
_{34}^{-}\rangle$, $|\Phi _{34}^{+}\rangle$, $|\Phi
_{34}^{-}\rangle$ after the unitary operations $u_0,u_1,u_2,u_3$
on anyone photon of the pair, respectively, that is, $u_0 |\Psi
_{34}^{+}\rangle=|\Psi _{34}^{+}\rangle$, $u_1 |\Psi
_{34}^{+}\rangle=|\Psi _{34}^{-}\rangle$, $u_2 |\Psi
_{34}^{+}\rangle=|\Phi _{34}^{+}\rangle$ and $u_3 |\Psi
_{34}^{+}\rangle=|\Phi _{34}^{-}\rangle$. Assume that each of the
above four unitary operations corresponds to a two-bit encoding
respectively, i.e., $u _{0}$ to '00', $u _{1}$ to '01', $u _{2}$
to '10' and $u _{3}$ to '11'. Then, taking advantage of the
quantum entanglement swapping and the assumption of the two-bit
codings, a deterministic secure direct communication protocol can
be proposed. We show it later.

Let us turn to depict our communication protocol. The protocol is
illustrated in figure 1 and works as follows:

\begin{figure}
 \begin{center}
 \includegraphics[width=1.0\textwidth]{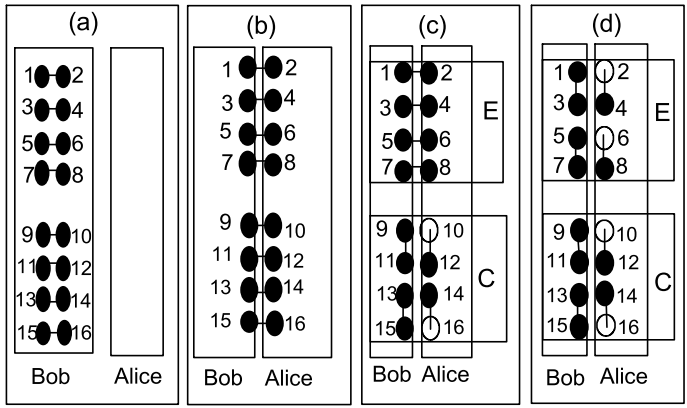}
 \vskip -3cm
 \caption{ Illustration of the present protocol. The black dots
 are photons. The hollow dot means a unitary operation has been
 performed on the photon. The line between two photons represents
 their entanglement. (a) Bob prepares the ordered $N$ EPR pairs.
 (b) Bob sends the travel photons to Alice. (c) The pairs are divided
 into two types of group. 'E' ('C') labels the encoding-decoding (checking) groups.
The checking procedure is performed. (d) The encoding and decoding
procedures are carried out.}
 \label{}
 \end{center}
\end{figure}

 (S1) Bob prepares a series of EPR pairs in $\Psi^{+}$ states (say,
the photon 1 and 2 pair, the photon 3 and 4 pair, etc). He takes
one photon from each pair, say, the photons 2,4,6,8, etc, to form
a string of photons in a regular sequence (say, the ordered string
of photons is '2468\dots'). He sends the ordered photon string to
Alice. In accordance with the ordering of the travel photons, Bob
stores the remainder photons  by way of two photons as a group,
i.e., the photons 1 and 3 as the group 1, the photons 5 and 7 as
the group 2, etc.

(S2) Alice confirms that she has received the travel photons. Also
in a regular sequence, she stores the arrived photons by way of
two photons as a group in terms of their coming orders, that is,
the photons 2 and 4 as the group 1, the photons 6 and 8 as the
group 2, etc.

(S3) Alice chooses randomly some photon groups as
encoding-decoding groups (say, the groups 1, 2, etc) for her later
two-bit encodings via local unitary $u$ operations. The remainder
photon groups are taken as checking groups. Alice first does
unitary $u$ operations on one photon of each checking group and
then performs the Bell measurements on these checking groups.

(S4) Alice publicly announces her exact unitary $u$ operation, her
Bell measurement result and the group order for each checking
group.

(S5) Bob performs the Bell measurements on his corresponding
photon groups whose orders are same with those publicly announced
by Alice. In the case of the group counterpart (i.e., the order of
Bob's group is same with the order of Alice's group), Bob compares
his measurement result with Alice's measurement result. If Bob
finds that each of Alice's measurement results is correlated with
his corresponding measurement result, he publicly tells Alice
there is no Eve in the line and the communication continues to
(S6). Otherwise, Bob publicly tells Alice that Eve is in the line.
The communication is aborted.

(S6) In accord with the encoding-decoding group ordering, Alice
performs her two-bit encodings via local unitary $u$ operations on
the encoding-decoding groups according to her bit strings (say,
'0110\dots') needed to be transmitted this time. For instances,
$u_1$ operation on one photon of the group 1 to encode '01', $u_2$
operation on one photon of the group 2 to encode '10', etc. After
the unitary $u$ operations, Alice performs the Bell states
measurements on all the encoding-decoding groups.

(S7) Alice publicly announces her Bell measurement result and the
encoding-decoding group order for each encoding-decoding photon
group.

(S8) Bob measures his unmeasured photon groups in the Bell states
after Alice's public announcement in (S7). After he knows each of
Alice's Bell measurement results with the group order and his Bell
measurement results with group orders, he can conclude the exact
unitary $u$ operations performed by Alice on each
encoding-decoding photon group, alternatively, he can extract the
two-bit encodings (cf. Table 1). In accordance with the photon
group ordering, he can get the bit string (i.e., the message). For
example, in (S6) since Alice had performed the $u_1$ operation on
one photon of the group 1 to encode '01', the initial state is
then changed into $\Psi_{12}^{+}\otimes \Psi_{34}^{-}$ (or
$\Psi_{12}^{-}\otimes \Psi_{34}^{+}$, which is equivalent to
$\Psi_{12}^{+}\otimes \Psi_{34}^{-}$ due to the symmetry) from the
state $\Psi_{12}^{+}\otimes \Psi_{34}^{+}$. If Alice obtains
$\Phi_{24}^{+}$ by her Bell measurements on her group 1, after
Alice's Bell measurement Bob should get $\Phi_{13}^{+}$ by his
Bell measurements on his group 1. Since Bob can know Alice's Bell
measurement result by her public announcement and his result by
his Bell measurement, he can conclude that Alice has performed the
operation $u _1$ and therefore extract the bits $(01)$ (see Table
1). Similarly, he can extract other bits, then he can get the bit
string '0110\dots'.

\begin{minipage}{340pt}
\begin{center}
\vskip 1cm Table 1.  Corresponding relations among the unitary $u$
operations (i.e., the encoding bits), the initial states, Bob's
and Alice's Bell measurement results.
\\
\begin{tabular}{cccc}  \hline
$u _0 (00)$ & $u _1 (01)$ & $u _2 (10)$ & $u _3 (11)$
\\ \hline
$\Psi_{12}^{+}\otimes \Psi_{34}^{+}$& $\Psi_{12}^{+}\otimes
\Psi_{34}^{-}$ & $\Psi_{12}^{+}\otimes \Phi_{34}^{+}$
&$\Psi_{12}^{+}\otimes \Phi_{34}^{-}$
\\ \hline
 $\left\{ \Phi
_{13}^{+},\Phi _{24}^{+}\right\} $ &
 $\left\{ \Psi _{13}^{-},\Psi
_{24}^{+}\right\} $ & $\left\{ \Psi _{13}^{+},\Phi
_{24}^{+}\right\} $ & $\left\{ \Psi _{13}^{+},\Phi _{24}^{-}\right\} $ \\
$\left\{ \Psi _{13}^{-},\Psi _{24}^{-}\right\} $ &  $\left\{ \Phi
_{13}^{+},\Phi _{24}^{-}\right\} $ &$\left\{ \Psi
_{13}^{-},\Phi _{24}^{-}\right\} $ & $\left\{ \Psi _{13}^{-},\Phi _{24}^{+}\right\} $ \\
$\left\{ \Psi _{13}^{+},\Psi _{24}^{+}\right\} $ & $\left\{ \Phi
_{13}^{-},\Phi _{24}^{+}\right\} $ & $\left\{ \Phi
_{13}^{+},\Psi _{24}^{+}\right\} $ & $\left\{ \Phi _{13}^{+},\Psi _{24}^{-}\right\} $ \\
$\left\{ \Phi _{13}^{-},\Phi _{24}^{-}\right\} $ &  $\left\{ \Psi
_{13}^{+},\Psi _{24}^{-}\right\} $ &$\left\{ \Phi _{13}^{-}, \Psi
_{24}^{-}\right\} $ & $\left\{ \Phi _{13}^{-},\Psi
_{24}^{+}\right\} $ \\ \hline
\end{tabular} \\
\end{center}
\end{minipage}
\vskip 1cm

Till now we have proposed a deterministic direct communication
protocol by utilizing the quantum entanglement swapping and local
unitary operations. To investigate the security of this
communication protocol, let us further consider some attack
scenarios by Eve.

(A) Let us consider the intercept-measure-resend attacks by Eve.
In this attack, Eve intercepts the ordered photon string
'2468\dots'. She classifies the travel photons as Alice does in
(S4), i.e., the photons 2 and 4 as the group 1, the photons 6 and
8 as the group 2, and so on. Then she performs her Bell
measurements on each group. After her measurements, she resends
the ordered string '2468\dots' to Alice. However, because of Eve's
direct measurement, the quantum entanglement swapping can not be
realized anymore. For examples, suppose Eve gets $\Phi_{24}^{+}$
by her Bell measurement on the photons 2 and 4, then Bob will
always get $\Phi_{13}^{+}$  by his Bell measurement on the
remainder photons 1 and 3. When Alice receives the photons 2 and 4
on which Bell measurement has been performed by Eve, she performs
first her $u_1$ ($u_2, u_3$) operations on the checking groups and
then the Bell measurements. After Alice publicly announces her
exact operation $u_0$ ($u_1$, $u_2, u_3$) and her Bell measurement
result $\Phi_{24}^{+}$ ($\Phi_{24}^{-}$, $\Psi_{24}^{+}$,
$\Psi_{24}^{-}$), Bob performs his Bell measurement on the
remainder photons 1 and 3 and gets inevitably $\Phi_{13}^{+}$.
Then in total Bob can find that his Bell measurement results do
not correlate with Alice's corresponding Bell measurement results
with possibility of $3/4$ corresponding to the $u_1$, $u_2$ and
$u_3$ operations but the occasional correlation with possibility
of $1/4$ corresponding to the $u_0$ operations. By the way, when
Alice performs her $u$ operations on the checking groups, she can
not exclude the $u_0$ operation to deterministically detect Eve's
attacks for Eve can also do the $u$ operations on the checking
groups. However, if the number of checking groups is large enough,
Bob can conclude that Eve is in the line because of the detection
possibility at the level high up to $3/4$. In addition, Eve can
get no information from Alice, because before Alice begins her
encoding, Eve is found in the line and accordingly the
communication is aborted. Therefore, the present protocol is
secure against the direct measurement attack by Eve.

(B) Let us consider the intercept-replace attacks by Eve. Assume
that Eve prepares some EPR pairs (i.e., the photon $1'$ and $2'$
pair, the photon $3'$ and $4'$ pair, etc) in the same states as
Bob prepares, i.e., in the states $\Psi ^{+}$. When Bob sends the
ordered photon string '2468\dots' to Alice, Eve intercepts the
photon sting and replaces it by her ordered photon string
'$2'4'6'8'\dots$' taken from her EPR pairs. After this Eve has two
choices, that is, either after or before Alice's Bell measurements
on the checking groups, she performs her Bell measurements on her
photons. Let us first consider Eve's first choice. In this case,
due to the replacement of the travel photons, the quantum
entanglement swapping can not be realized anymore, therefore there
are no inevitable correlations between Alice' Bell measurement
results on the photons $2'$ and $4'$ and Bob's Bell measurement
results on the photons $1$ and $3$. Hence, it is easy for Alice
and Bob to find that Eve is in the line in terms of their joint
actions on the checking groups. This is very similar to the
detection on Eve in (A). So in this case the present protocol is
also secure. Let us turn to consider Eve's second choice, i.e.,
Eve performs her Bell measurements at first. In this case, it is
obvious that Eve can not get any information from Alice's encoding
by her prior Bell measurements. In the following let us
investigate the detection on Eve. If Eve performs her Bell
measurements on the photon $1'$ and $3'$ pair and the photon 2 and
4 pair, then the situation turns to her first choice. Therefore,
she should carry out her Bell measurements on the photon $1'$ and
2 pair and the the photon $3'$ and 4 pair. Incidentally, her Bell
measurements on the photon $1'$ and 4 pair and the the photon $3'$
and 2 pair are equivalent to her Bell measurements on the photon
$1'$ and 2 pair and the the photon $3'$ and 4 pair due to the
symmetry. The initial state of the system including Eve's EPR
pairs is
\begin{eqnarray}
&& [\Psi_{12}^{+}\otimes
\Psi_{34}^{+}]\otimes[\Psi_{1'2'}^{+}\otimes \Psi_{3'4'}^{+}] \nonumber \\
&=& \frac{1}{4}(|\Psi _{13}^{+}\rangle |\Psi _{24}^{+}\rangle
-|\Psi _{13}^{-}\rangle|\Psi _{24}^{-}\rangle+|\Phi
_{13}^{+}\rangle|\Phi _{24}^{+}\rangle-|\Phi _{13}^{-}\rangle|\Phi
_{24}^{-}\rangle)  \nonumber \\ &&\otimes (|\Psi
_{1'3'}^{+}\rangle |\Psi _{2'4'}^{+}\rangle -|\Psi
_{1'3'}^{-}\rangle|\Psi _{2'4'}^{-}\rangle+|\Phi
_{1'3'}^{+}\rangle|\Phi
_{2'4'}^{+}\rangle-|\Phi _{1'3'}^{-}\rangle|\Phi _{2'4'}^{-}\rangle)  \nonumber \\
&=& \frac{1}{4}[ |\Psi _{13}^{+}\rangle  |\Psi
_{2'4'}^{+}\rangle|\Psi _{24}^{+}\rangle|\Psi _{1'3'}^{+}\rangle
+|\Psi _{13}^{-}\rangle|\Psi _{2'4'}^{-}\rangle |\Psi
_{24}^{-}\rangle|\Psi _{1'3'}^{-}\rangle  \nonumber \\ && +|\Phi
_{13}^{+}\rangle|\Phi _{2'4'}^{+}\rangle|\Phi
_{24}^{+}\rangle|\Phi _{1'3'}^{+}\rangle+|\Phi
_{13}^{-}\rangle|\Phi _{2'4'}^{-}\rangle|\Phi
_{24}^{-}\rangle|\Phi _{1'3'}^{-}\rangle]  \nonumber \\
&+& \frac{1}{4}[|\Psi _{13}^{+}\rangle |\Psi _{24}^{+}\rangle
(-|\Psi _{1'3'}^{-}\rangle|\Psi _{2'4'}^{-}\rangle+|\Phi
_{1'3'}^{+}\rangle|\Phi
_{2'4'}^{+}\rangle-|\Phi _{1'3'}^{-}\rangle|\Phi _{2'4'}^{-}\rangle)  \nonumber \\
&& -|\Psi _{13}^{-}\rangle|\Psi _{24}^{-}\rangle(|\Psi
_{1'3'}^{+}\rangle |\Psi _{2'4'}^{+}\rangle +|\Phi
_{1'3'}^{+}\rangle|\Phi
_{2'4'}^{+}\rangle-|\Phi _{1'3'}^{-}\rangle|\Phi _{2'4'}^{-}\rangle)  \nonumber \\
&& +|\Phi _{13}^{+}\rangle|\Phi _{24}^{+}\rangle(|\Psi
_{1'3'}^{+}\rangle |\Psi _{2'4'}^{+}\rangle -|\Psi
_{1'3'}^{-}\rangle|\Psi _{2'4'}^{-}\rangle-|\Phi
_{1'3'}^{-}\rangle|\Phi _{2'4'}^{-}\rangle)
 \nonumber \\ && -|\Phi _{13}^{-}\rangle|\Phi _{24}^{-}\rangle
 (|\Psi _{1'3'}^{+}\rangle |\Psi _{2'4'}^{+}\rangle
-|\Psi _{1'3'}^{-}\rangle|\Psi _{2'4'}^{-}\rangle+|\Phi
_{1'3'}^{+}\rangle|\Phi _{2'4'}^{+}\rangle)].
\end{eqnarray}
So after Eve's Bell measurements on the photon $1'$ and 2 pair and
the photon $3'$ and 4 pair, when Alice and Bob perform their joint
actions on the checking group, in each checking group the
detection possibility on Eve is also at the level high up to $3/4$
due to Eve's attacks. Therefore, when the number of the checking
group is big enough, then the present protocol is also secure
under Eve's such attacks.

(C) Let us consider the adding-ancilla attacks by Eve. Suppose Eve
can take advantage of CNOT gate operations as used in [25] to add
an ancilla to each travel photon. Then the initial state including
the ancillas is
\begin{eqnarray}
&& C_{22'}\Psi_{12}^{+}|0\rangle_{2'}\otimes C_{44'}
\Psi_{34}^{+}|0\rangle_{4'} \nonumber \\ &=&
\frac{1}{2}(|0\rangle_1|1\rangle_2|1\rangle_{2'}+
|1\rangle_1|0\rangle_2|0\rangle_{2'}) \otimes
(|0\rangle_3|1\rangle_4|1\rangle_{4'}+
|1\rangle_3|0\rangle_4|0\rangle_{4'}) \nonumber \\
&=& \frac{\sqrt{2}}{4}[(|\Psi _{13}^{+}\rangle |\Psi
_{24}^{+}\rangle -|\Psi _{13}^{-}\rangle|\Psi
_{24}^{-}\rangle)|\Psi _{2'4'}^{+}\rangle +(|\Phi
_{13}^{+}\rangle|\Phi _{24}^{+}\rangle-|\Phi _{13}^{-}\rangle|\Phi
_{24}^{-}\rangle)|\Phi _{2'4'}^{+}\rangle \nonumber \\
&& + (|\Psi _{13}^{+}\rangle |\Psi _{24}^{-}\rangle -|\Psi
_{13}^{-}\rangle|\Psi _{24}^{+}\rangle)|\Psi _{2'4'}^{-}\rangle
+(|\Phi _{13}^{+}\rangle|\Phi _{24}^{-}\rangle-|\Phi
_{13}^{-}\rangle|\Phi _{24}^{+}\rangle)|\Phi _{2'4'}^{-}\rangle],
\end{eqnarray}
where $C$ stands for the CNOT operation. From this equation, one
can find that Eve can not get any information form Alice's
encoding under this kind of attacks. Let us further investigate
the detection possibility on Eve. First, assume that Eve performs
her Bell measurement on her two ancilla photons before Alice and
Bob's joint actions. When Eve gets $|\Psi _{2'4'}^{-}\rangle$ or
$|\Phi _{2'4'}^{-}\rangle$), she performs the local $\sigma_1$
operation on one of the two travel photons. In this case, Eve can
successfully avoid the detection on her. However, her attack does
not affect the communication between Alice and Bob at all.
Secondly, assume that Eve does noting but adding the ancilla
photons. In this case, for each checking group Alice and Bob can
find Eve with possibility of $1/2$ in terms of their joint
actions. So when the number of the checking groups is big enough,
then Eve can be found in the line. So the present protocol is also
secure against such attacks.

To summarize, under the above attack scenarios to our best
knowledge, the present protocol is secure. So in this paper, we
have proposed a deterministic secure direct quantum communication
protocol by using the entanglement swapping and local unitary
operations. Incidentally, an alternative scheme of the present
protocol can be achieved, i.e., different from the fact that it is
always Bob who prepares the entangled photon pairs and sends the
travel photons to Alice in the present scheme, in the alternative
scheme both Alice and Bob prepare the entangled photon pair and
they send one photon of the entangled pair to each other.

This work is supported by the National Natural Science Foundation
of China under Grant No. 10304022.

\newpage

\noindent [1] C. H. Bennett and G. Brassard, in {\it Proceedings
of the IEEE International Conference on Computers, Systems and
Signal Processings, Bangalore, India} (IEEE, New York, 1984),
p175.

\noindent[2] A. K. Ekert, Phys. Rev. Lett. {\bf67}, 661 (1991).

\noindent[3] C. H. Bennett, Phys. Rev. Lett. {\bf68}, 3121
 (1992).

\noindent[4] C. H. Bennett, G. Brassard, and N.D. Mermin, Phys.
Rev. Lett. {\bf68}, 557(1992).

\noindent[5] L. Goldenberg and L. Vaidman, Phys. Rev. Lett.
{\bf75}, 1239  (1995).

\noindent[6] B. Huttner, N. Imoto, N. Gisin, and T. Mor, Phys.
Rev. A {\bf51}, 1863 (1995).

\noindent [7] M. Koashi and N. Imoto, Phys. Rev. Lett. {\bf79},
2383 (1997).

\noindent[8] W. Y. Hwang, I. G. Koh, and Y. D. Han, Phys. Lett. A
{\bf244}, 489 (1998).

\noindent[9] P. Xue, C. F. Li, and G. C. Guo,  Phys. Rev. A
{\bf65}, 022317 (2002).

\noindent[10] S. J. D. Phoenix, S. M. Barnett, P. D. Townsend, and
K. J. Blow, J. Mod. Opt. {\bf42}, 1155 (1995).

\noindent[11] H. Bechmann-Pasquinucci and N. Gisin, Phys. Rev. A
{\bf59}, 4238 (1999).

\noindent[12] A. Cabello, Phys. Rev. A {\bf61},052312 (2000);
{\bf64}, 024301 (2001).

\noindent[13] A. Cabello, Phys. Rev. Lett. {\bf85}, 5635 (2000).

\noindent[14] G. P. Guo, C. F. Li, B. S. Shi, J. Li, and G. C.
Guo, Phys. Rev. A {\bf64}, 042301 (2001).

\noindent[15] A. Beige, B. G. Englert, C. Kurtsiefer, and
H.Weinfurter, Acta Phys. Pol. A {\bf101}, 357 (2002).

\noindent[16] Kim Bostrom and Timo Felbinger, Phys. Rev. Lett.
{\bf89}, 187902 (2002).

\noindent[17] G. L. Long and X. S. Liu, Phys. Rev. A {\bf65},
032302 (2002).

\noindent[18] F. G. Deng and G. L. Long, Phys. Rev. A {\bf68},
042315 (2003).

\noindent[19] F. G. Deng, G. L. Long, and X. S. Liu, Phys. Rev. A
{\bf68}, 042317 (2003).

\noindent[20] Daegene Song, Phys. Rev. A {\bf69}, 034301(2004).

\noindent [21] M. Zukowski, A. Zeilinger, M. A. Horne, and  A. K.
Ekert, Phys. Rev. Lett. {\bf71}, 4287 (1993).

\noindent [22] S. Bose, V. Vedral, and P. L. Knight, Phys. Rev. A
{\bf57}, 822(1998).

\noindent [23] L. Hardy and D. Song, Phys. Rev. A {\bf62},
052315(2000).

\noindent [24] J. W. Pan, M. Daniell, S. Gasparoni, G. Weihs, and
A. Zeilinger, Phys. Rev. Lett. {\bf86}, 4435 (2001).

\noindent [25] A. Wojcik, Phys. Rev. Lett. {\bf90}, 157901 (2003).

\end{document}